\documentclass[aps,prl,showpacs,twocolumn,superscriptaddress]{revtex4}
\usepackage{amsmath}
\usepackage{amsfonts}
\usepackage{graphicx}
\usepackage{subfigure}

\begin{document}

\title{Networks of quantum nanorings: programmable spintronic devices}
\author{P\'{e}ter F\"{o}ldi}
\affiliation{Department of Theoretical Physics, University of Szeged, Tisza Lajos k\"{o}r%
\'{u}t 84, H-6720 Szeged, Hungary}
\author{Orsolya K\'{a}lm\'{a}n}
\affiliation{Department of Quantum Optics and Quantum Information, Research Institute for Solid
State Physics and Optics,Hungarian Academy of Sciences, Konkoly-Thege Mikl\'{o}s \'{u}t 29-33, H-1121 Budapest, Hungary}
\author{Mih\'{a}ly G. Benedict}
\email{benedict@physx.u-szeged.hu}
\affiliation{Department of Theoretical Physics, University of Szeged, Tisza Lajos k\"{o}r%
\'{u}t 84, H-6720 Szeged, Hungary}
\author{F. M. Peeters}\email{francois.peeters@ua.ac.be}
\affiliation{Departement Fysica, Universiteit Antwerpen, Groenenborgerlaan
171, B-2020 Antwerpen, Belgium}

\begin{abstract}
An array of quantum rings with local (ring by ring) modulation of the spin orbit interaction (SOI)
can lead to novel effects in spin state transformation of electrons.
It is shown that already small ($3\times 3,$ $5 \times 5$) networks are remarkably versatile from this point
of view: Working in a given network geometry, the input current can be directed
to any of the output ports, simply by changing the SOI strengths by external gate voltages. Additionally,
the same network with different SOI strengths can be completely analogous to the Stern-Gerlach device, exhibiting
spatial-spin entanglement.
\end{abstract}

\maketitle

Using the spin degree of freedom in information processing applications
has become an important and rapidly developing field. For most of the
spintronic devices, spin is a classical resource, logical states are "up" and
"down" with respect to certain quantization directions, but their superpositions
play no role. On the other hand, the electron spin degree of freedom can
also be considered one of the prospective carriers of qubits, the fundamental
units in quantum information processing \cite{ALS08}. To realize
this aim, however, requires to perform basic spin operations such as the production of spin-polarized
carriers and spin rotation. Here we propose a device
that can serve multiple purposes including the delivery of
spin-polarized currents and rotating the spin direction, but it can also direct the input
current into a given output port in a spin independent way. The significance of
these results is related to the flexibility of the device: the same geometry
provides qualitatively different transport properties in such a way that
parameters are being varied in an experimentally achievable range.

We calculate the spin transport properties
of two-dimensional rectangular arrays of nanoscale quantum rings,
which can be fabricated from e.g. InAlAs/InGaAs based heterostructures \cite{KNAT02} or
HgTe/HgCdTe quantum wells \cite{KTHS06},
where Rashba-type \cite{R60} spin-orbit interaction (SOI) is present.
This effect, which is essentially the same as the one which causes the fine structure in atomic spectra,
results in spin precession for electrons moving in a semiconductor.
It has already been demonstrated in experiments that the strength of this type of SOI
can be controlled by external gate voltages \cite{NATE97,G00} in the range of a few Volts.
We propose devices in which the local manipulation of the SOI strength leads to effects
which could be used in various practical spintronic applications. We focus on narrow rings in the ballistic
(coherent) regime, where a one dimensional model provides appropriate description.

\begin{figure}[tbh]
\begin{center}
\includegraphics*[width=8.5cm]{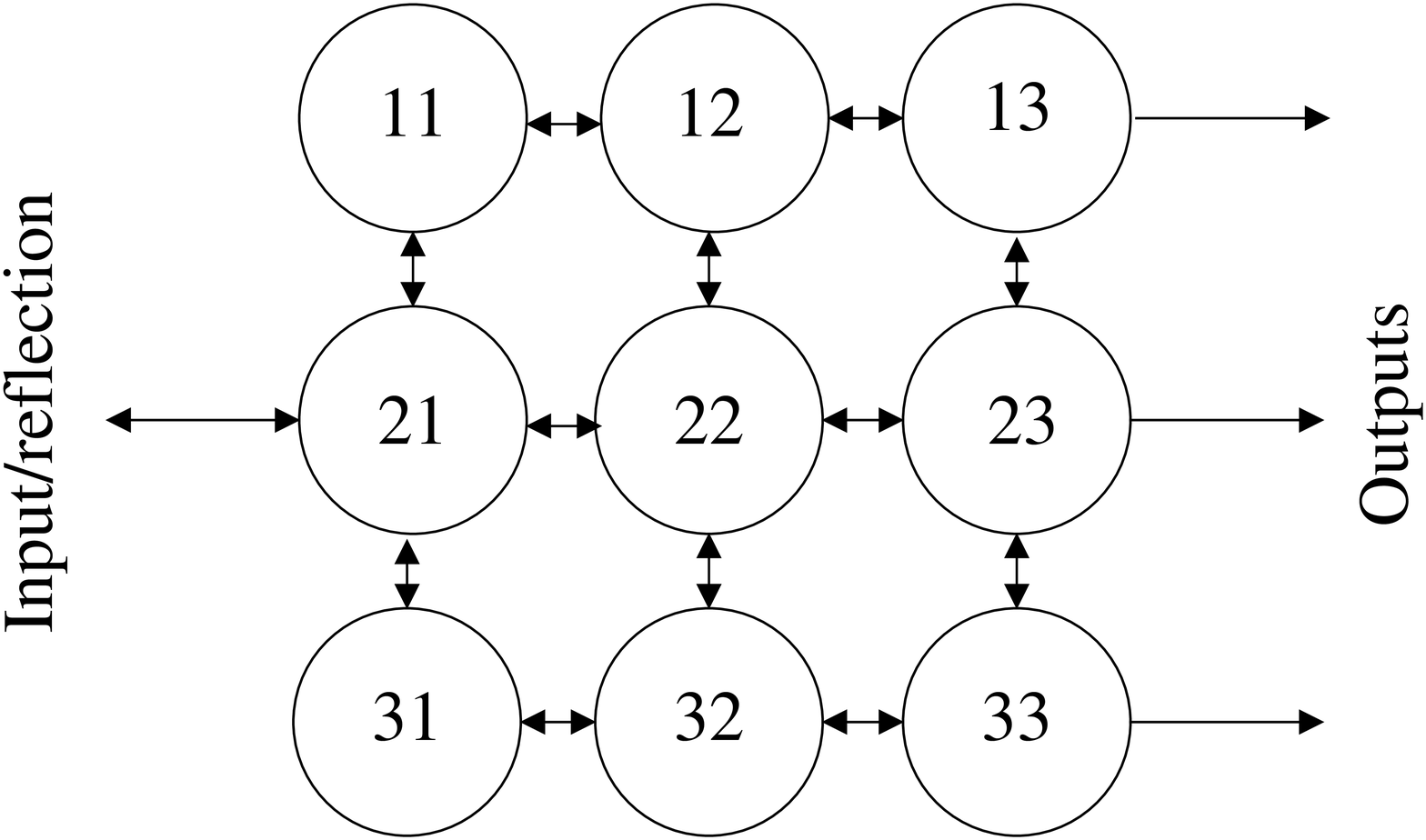}
\end{center}
\caption{ The geometry of a $3 \times 3$ ring array. The arrows indicate the possible directions of
the currents. Note that on the output side there are only outgoing solutions, while the possible
reflection of the incoming spinor valued wave function is also taken into account.}
\label{networkfig}
\end{figure}

The geometries we are considering (namely $3\times 3$\ and
$5\times 5$\ ring arrays, see Fig.~\ref{networkfig}) have already appeared in a recent
experiment \cite{BKSN06} with rings of 100 nm thickness, and the case of uniform SOI strength has also been investigated
theoretically \cite{ZW07,KFBP08b}. We note that similarly to Ref.~\cite{BKSN06}, the rings we consider here touch
each other, thus the lines between them shown in the figures serve only visualization purposes.
Considering wires of finite length would not pose any
problem from the theoretical point of view, and this may be required
technically to realize the ring by ring modulation of the SOI strength.

Our method uses a building block principle, first we solve the spin dependent scattering problem for
single rings, which is possible analytically, and then we use these results to derive
the properties of two-dimensional arrays. We note that spin transformation properties of single quantum rings in the presence
of Rashba SOI \cite%
{NMT99,FHR01,YPS03,MPV04,FMBP05,SN05,KMGA05,SP05,FKBP06,CHR07,BO07,KFBP08}, and linear chains of such
rings \cite{MVP05} have already been investigated, but the current results
cannot be simply deduced from the previous ones, as they are
related to complex quantum mechanical interference phenomena involving the entire
array, not just smaller parts of it. At the junctions between the rings, the solutions have to be fitted,
and the corresponding large number of coupled linear equations are solved numerically.

\bigskip

First we consider a single, narrow quantum ring of radius $a$ located in the $xy$ plane.
The relevant dimensionless Hamiltonian reads \cite{MMK02}%
\begin{equation*}
\tilde{H}=\left( -i\frac{\partial }{\partial \varphi }+
\frac{\omega _{SO}}{2\Omega }\sigma _{r}\right) ^{2}-\frac{\omega _{SO}^{2}%
}{4\Omega ^{2}},  \label{Ham}
\end{equation*}%
where $\varphi $ is the azimuthal angle of a point on the ring,
and $\omega _{SO}=\alpha /\hbar a$ is the frequency associated with the
spin-orbit interaction, which can be changed by an external gate voltage that tunes the
value of $\alpha$. $\hbar\Omega =\hbar^{2} /2m^{\ast }a^{2}$ gives
the kinetic energy with \thinspace $m^{\ast }$ being the effective mass of
the electron, and the radial spin operator in units of $\hbar$ is given by
$\sigma _{r}/2=(\sigma _{x}\cos\varphi +\sigma _{y}\sin\varphi)/2$.
The energy eigenvalues and the corresponding
eigenstates of this Hamiltonian can be calculated analytically \cite{MPV04,FMBP05}. For a ring with
leads attached to it, the spectrum is continuous, all positive energies can appear,
and they are fourfold degenerate. This degeneracy is related to i) two possible eigenspinor
orientations and ii) to the two possible (clockwise and anticlockwise) directions in which
currents can flow.
The state of the incoming electron is considered to be a
plane wave with wave number $k$. By energy conservation its energy
(given by $E=\hbar ^{2}k^{2}/2m^{\ast}$) determines the solutions in the rings.
At the junctions (between the incoming lead and ring 21, the outgoing leads and
rings 13, 23, 33, as well as between different rings, see Fig.~\ref{networkfig})
Griffith's boundary conditions \cite{G53}
are applied, i.e., the net spin current density at a certain junction has to vanish,
and we also require the continuity of the spinor valued wave functions. If we finally
specify that there are no incoming electrons on the output side, an unambiguous solution for
the spin-dependent scattering problem can be found \cite{KFBP08b}.
\bigskip

In order to best utilize the spin transformation potential of the device, we consider the case
when the SOI strengths, instead of being uniform, are modulated locally, ring by ring.
In other words, $\omega _{SO}/\Omega$ is constant within a given ring, but can be different for
different rings. These conditions lead to various spin-dependent
transport properties, here we consider two specific examples that can have practical applications
as well. We focus on small networks: although larger arrays offer even more
possibilities, the condition of sustaining quantum coherence is more difficult
to maintain experimentally for larger distances.

\begin{figure}[tbh]
\begin{center}
\includegraphics*[width=8.5cm]{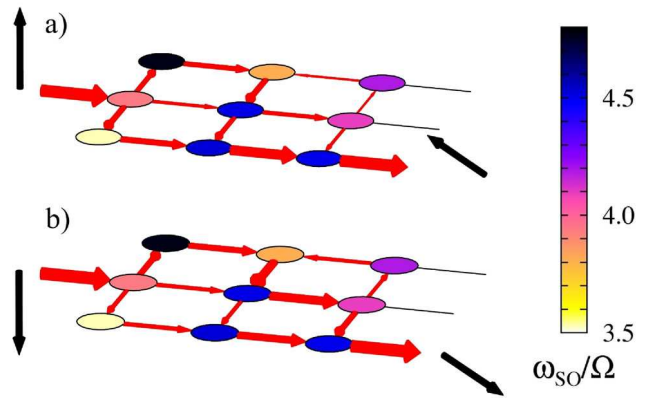}
\end{center}
\caption{Directing the input current to a chosen output port for $ka=20.67.$
Upper (lower) panel shows the case of spin-up (spin-down) input (black arrows).
Red arrows visualize the currents flowing between the rings, and the local SOI strengths
are indicated by color coding.}
\label{swfig}
\end{figure}

\bigskip

We begin the presentation of the results with an interesting switching effect shown in Fig.~\ref{swfig}:
the current in the network can be directed by SOI modulation to a given output port. This kind of device can be
made practically reflectionless, and the probability for an electron to leave the device through
a lead other than the distinguished one is less than 1\%. The most remarkable point concerning
this result is that it is spin independent on the level of the conductance: The output port is
always the same, \emph{regardless} the input spin direction. Spin-dependence becomes apparent when we
calculate the direction of the spin of the output electron: As the arrows at the different
ports show in Fig.~\ref{swfig}, the output spin states are orthogonal for the $z$ direction
spin-up and spin-down inputs. (We found that in the case of a $5 \times 5$ array even the orientation of
the output spins can be controlled.) Additionally, using different SOI strengths in the same $3 \times 3$
array as shown in Fig.~\ref{swfig}, the current can be directed to any of the output ports.

\begin{figure}[tbh]
\begin{center}
\includegraphics*[width=8.5cm]{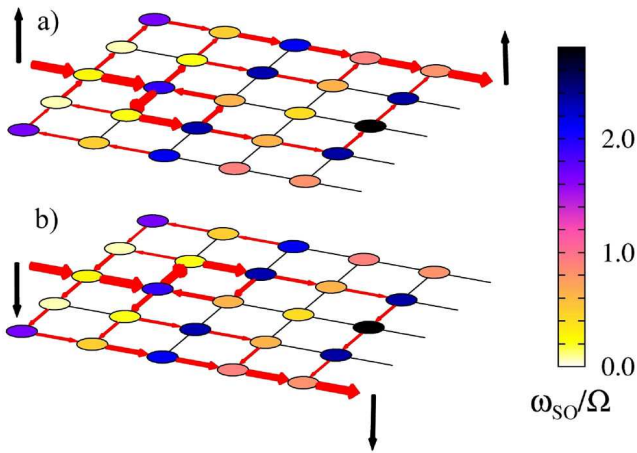}
\end{center}
\caption{$5 \times 5$ network (with $ka=24.9$) acting as a spintronic analogue
of the Stern-Gerlach device.  The meaning of the arrows is the same as in Fig.~\ref{swfig}.}
\label{sgfig}
\end{figure}

The effect visualized by Fig.~\ref{swfig}
raises the question whether it is possible to realize a device that spatially separates the orthogonal inputs.
According to our calculations, the same $3 \times 3$ array can also lead to this kind of
separation effect. For slightly larger arrays, the orientation of the outputs can be chosen to
coincide with the input. As it is seen in Fig.~\ref{sgfig}, in a $5 \times 5$ network with appropriate
parameters, if the input is one of the eigenstates of $\sigma_z,$ the output will have the same
spin direction at a certain output port (with probability
higher than 99\%). The orthogonal input is directed towards a different output port,
with its final direction being the same as the initial one. According to the superposition
principle, the device with these parameters will always produce orthogonal $\sigma_z$
eigenstates at the two distinguished outputs. The amplitudes of the different output states
is the same as that of the corresponding $\sigma_z$ eigenstates in the input. On the other hand,
when the input is unpolarized, i.e., it is an incoherent sum of spin-up and spin-down states, a
polarizing effect appears: the outputs will be the same $\sigma_z$ eigenstates as before.
Let us note that although a similar polarizing effect is already present in
the case of a single ring with one input and two output ports \cite{FKBP06,KFBP08}, the current results are remarkable
as they describe a device that can be considered as a more complete spintronic analogue of the
Stern-Gerlach apparatus, including not only the spatial separation, but also the orthogonality
of the output spinors. This represents a typical example of entanglement (intertwining) of different
-- in our case spatial and spin -- degrees of freedom \cite{KFB06}.

Let us finally note that although single rings already have significant spin transformation properties,
they are, in view of the current results, limited in the sense that a fixed geometry (ring size and
positions of the attached leads) usually can serve only for one given purpose.
Also the well-established results related to single quantum rings cannot easily be
extended for the construction of networks aimed to be multipurpose spintronic devices.
In contrast, the network considered here is a flexible device, the transformation
properties are tunable during operation by gating the different rings.
We also stress that this is a nontrivial extension of connected single rings,
as the spin dependent quantum mechanical interference that leads to a specific behavior is a global phenomenon,
it involves the array as a whole. Considering the size of the network, it is clear that
due to the increasing number of possible paths that can interfere, larger arrays have higher spin
transformational potential -- provided they can be considered to be ballistic. By representing an optimal
compromise between size and spin transformational properties, $3 \times 3$ networks can be the most promising
for practical applications.

\bigskip
In conclusion, we investigated arrays of quantum rings with Rashba-type spin orbit interaction (SOI).
It was shown that by locally tuning the SOI strengths in a given ring network, various important
spintronic devices can be realized.

\section*{Acknowledgments}
This work was supported by the Flemish-Hungarian Bilateral
Programme, the Flemish Science Foundation (FWO-Vl), the Belgian Science Policy
and the Hungarian Scientific Research Fund (OTKA) under Contracts Nos.~T48888,
M36803, M045596. P.F.~was supported by a J.~Bolyai grant of the
Hungarian Academy of Sciences.


\end{document}